\documentclass[12pt]{article}
\usepackage{amssymb,amscd,array}
\catcode `\@=11
\@addtoreset{equation}{subsection}

\def\harr#1#2{\smash{\mathop{\hbox to .5in{\rightarrowfill}}
\limits^{\scriptstyle#1}_{\scriptstyle#2}}}

\def\harrl#1#2{\smash{\mathop{\hbox to .5in{\leftarrowfill}}
\limits^{\scriptstyle#1}_{\scriptstyle#2}}}

\def\qed{\blacksquare}
\newcommand{\be}{\begin{equation}}
\newcommand{\ee}{\end{equation}}
\newcommand{\bea}{\begin{eqnarray}}
\newcommand{\eea}{\end{eqnarray}}
\newcommand{\R}{\mathbb{R}}
\newcommand{\N}{\mathbb{N}}
\newcommand{\C}{\mathbb{C}}
\newtheorem{thm}{Theorem}[section]
\newtheorem{rem}[thm]{Remark}

\begin{document}
\begin{titlepage}
\begin{center}
{\bf \Large{The Renormalization of the Non-Abelian Gauge Theories 
in the Causal Approach\\}}
\end{center}
\vskip 1.0truecm
\centerline{D. R. Grigore
\footnote{e-mail: grigore@theor1.theory.nipne.ro, grigore@theory.nipne.ro}}
\vskip5mm
\centerline{Dept. of Theor. Phys., Inst. Atomic Phys.}
\centerline{Bucharest-M\u agurele, P. O. Box MG 6, ROM\^ANIA}
\vskip 2cm
\bigskip \nopagebreak
\begin{abstract}
\noindent
We consider the gauge invariance of the standard Yang-Mills model in the
framework of the causal approach of Epstein-Glaser and Scharf and determine the
generic form of the anomalies. The method used is based Epstein-Glaser 
approach to renormalization theory. In the case of quantum electrodynamics we
obtain quite easily the absence of anomalies in all orders.
\end{abstract}
\end{titlepage}

\section{Introduction}

The causal approach to renormalization theory of by Epstein and Glaser
\cite{EG1}, \cite{Gl} had produced important simplification of the
renormalization theory at the purely conceptual level as well as to the
computational aspects. This approach works for quantum electrodynamics
\cite{Sc1}, \cite{DF1}, \cite{qed} where it brings important simplifications of
the renormalizability proof. For Yang-Mills theories \cite{DHKS1},
\cite{DHKS2}, \cite{DHS2}, \cite{DHS3}, \cite{AS}, \cite{ASD3}, \cite{Du1}, 
\cite{Du2}, \cite{Hu1}-\cite{Hu4}, \cite {Kr1}, \cite{DS}, \cite{YM},
\cite{standard}, \cite{fermi}, \cite{Sc4} one can determine severe constraints
on the interaction Lagrangian (or in the language of the renormalization theory
- on the first order chronological product) from the condition of gauge
invariance.  Gravitation can be also analysed in this framework
\cite{gravity}, \cite{Gri1}, \cite{Gri2}, \cite{SW}, etc. Finally, the analysis
of scale invariance can be done \cite{scale}, \cite{Pr3}. One should stress the
fact that the Epstein-Glaser analysis uses exclusively the Bogoliubov axioms of
renormalization theory \cite{BS2} imposed on the scattering matrix: this is an
operator acting in the Hilbert space of the model, usually generated from the
vacuum by the quantum fields corresponding to the particles of the model. If
one considers the $S$-matrix as a perturbative expansion in the coupling
constant of the theory, one can translate these axioms on the chronological
products. Epstein-Glaser approach is a inductive procedure to construct the
chronological products in higher orders starting from the first-order of the
perturbation theory. For gauge theories one can construct a non-trivial
interaction only if one considers a larger Hilbert space generated by the
fields associated with the particles of the model and the ghost fields. The
condition of gauge invariance becomes in this framework the condition of
factorization of the $S$-matrix to the physical Hilbert space in the adiabatic
limit. To avoid infra-red problems one works with a formulation of this
factorization condition which corresponds to a formal adiabatic limit and it is
perfectly rigorously defined \cite{DHKS2}. The obstructions to the
implementation of the condition of gauge invariance are called {\it anomalies}.
The most famous is the Adler-Bell-Bardeen-Jackiw anomaly \cite{Ad}, \cite{Ba},
\cite{BJ}, \cite{Be} (see \cite{Ni1} for a review).

The classical analysis of the renormalizablity of Yang-Mills theories of
Becchi, Rouet, Stora and Tyutin \cite{BRS} is based on a different
combinatorial idea. Namely, one considers a perturbative expansion in Planck
constant
$\hbar$ 
which is equivalent, in Feynman graphs terminology, to a loop expansion. (The
rigorous connection between these two perturbation schemes has been recently
under investigation \cite{DF3}.) One can formulate the condition of gauge
invariance in terms of the generating functional for the one-particle
irreducible Feynman amplitudes; the $S$-matrix is then recovered using the
reduction formul\ae~ \cite{PR}. Presumably, both formulations lead to the same
$S$-matrix, up to finite renormalization, although this point is not firmly
established in the literature. The most difficult part is to prove that if
there are no anomalies in lower orders of perturbation theory, then the
anomalies are absent in higher orders. The main tool of the proof is the
consideration of the scale invariance properties of a quantum theory expressed
in the form of Callan-Symanzik equations \cite{Ca}, \cite{CCJ} and \cite{CJ}. A
mathematical analysis was developed in \cite{Si1} and \cite{Si2}, using the
quantum action principle \cite{Lo} (for a review see \cite{PR}). One should
stress the fact that in this approach one works with interaction fields which
can be defined as formal series in the coupling constant. The main observation
used in these references is the existence of anomalous dimensional behaviour of
the (interacting) fields with respect to dilations. Based on this analysis in
\cite{BBBC} (see also \cite{BRS} and \cite{PR}) it is showed that the ABBJ
anomaly can appear only in the order
$n = 3$ 
of the perturbation theory.  A analysis of the standard model based on this
approach can be found in \cite{Kraus}.

In \cite{scale} we have investigated scale anomaly from the point of view of
Epstein-Glaser causal approach based entirely on a perturbation scheme of
Bogoliubov based on an expansion in the coupling constant. We have found out
the surprising result that scale invariance does not restrict the presence of
the anomalies in higher orders of perturbation theory. So, from the point of
view of Bogoliubov axioms, the elimination of anomalies in higher orders of
perturbation theory is still an open question.

The purpose of this paper is to investigate the generic form of the anomalies
compatible with the restrictions following from covariance properties and
formal gauge invariance. Our strategy will be based exclusively on the
Epstein-Glaser construction of the chronological products for the free fields.
The r\^ole of Feynman graph combinatorics is completely eliminated in this
analysis. We will use in fact a reformulation of the Epstein-Glaser formalism
\cite{DF1} which gives a prescription for the construction of the chronological
products of the type
$
T(A_{1}(x_{1}),\cdots,A_{n}(x_{n}))
$
for any Wick polynomials
$
A_{1}(x_{1}),\cdots,A_{n}(x_{n}).
$
The main point is to formulate a proper induction hypothesis for the expression
$
d_{Q} T(A_{1}(x_{1}),\cdots,A_{n}(x_{n}))
$
where
$d_{Q}$
is the BRST operator\footnote{M. D\"utsch, private communication}. In fact, we
will see that it is necessary to make such a conjecture only for some special
cases of Wick polynomials. If
$T(x)$
is the interaction Lagrangian (i.e. the first order chronological product) one 
can prove the validity of some ``descent" equations of the type:
\be
d_{Q} T(x) = i \partial_{\mu}T^{\mu}(x), \quad
d_{Q} T^{\mu}(x) = i \partial_{\nu}T^{\mu\nu}(x), \quad, \dots, \quad
d_{Q} T^{\mu_{1},\dots,\mu_{p-1}}(x) = 
i \partial_{\mu_{p}}T^{\mu_{1},\dots,\mu_{p}}(x).
\ee

In the QED the procedure stops after the first step 
($p = 1$)
and in the Yang-Mills case after a two steps
($p = 2$).
In general, one can consider the case when the descent stops after a finite
number of steps. In this case one has to give a proper conjecture for
$
d_{Q} T(A_{1}(x_{1}),\cdots,A_{n}(x_{n})) 
$ 
only for 
$
A_{1}(x_{1}),\cdots,A_{n}(x_{n}) $ of the type $ T(x), \quad T^{\mu}(x),
\quad T^{\mu\nu}(x), \quad \dots.
$

The structure of this paper is the following one. In the next Section we make a
brief review of essential points concerning Epstein-Glaser resolution scheme of
Bogoliubov axioms and the standard model in the framework of the causal
approach. (For more details see \cite{qed} and \cite{standard}). We emphasize
that the main problem is to establish the factorization of the $S$-matrix to
the physical Hilbert space; in the formal adiabatic limit, this is the famous
condition of gauge invariance.  Translated in terms of Feynman amplitudes this
condition amounts, essentially, to the so-called Ward-Takahashi identities, or
- in the language of the Z\"urich group - the C-g identities. In the next
Section we give the inductive hypothesis for quantum electrodynamics and prove
that there are no anomalies. Next, we do the same thing for Yang-Mills theories
and determine the generic form of the anomalies.
\newpage

\section{Perturbation Theory in the Causal Approach\label{pert}}

\subsection{Bogoliubov Axioms}{\label{bogoliubov}}

We present the main ideas of perturbation theory following \cite{EG1} and
\cite{qed} to which we refer for more details. The $S$-matrix is formal series
of operator valued distributions:
\be
S({\bf g})=1 + \sum_{n=1}^\infty{i^{n}\over n!}\int_{\R^{4n}} 
dx_{1}\cdots dx_{n}\, T_{j_{1},\dots,j_{n}}(x_{1},\cdots, x_{n})
g_{j_{1}}(x_{1})\cdots g_{j_{n}}(x_{n}),
\label{S-matrix}
\ee
where
${\bf g} = \left( g_{j}(x)\right)_{j = 1, \dots P}$
is a vector-valued tempered test function and
$T_{j_{1},\dots,j_{n}}(x_{1},\cdots, x_{n})$
are operator-valued distributions acting in the Fock space of some collection
of free fields with a common dense domain of definition
$D_{0}$. 
The scalar product is denoted by
$(\cdot,\cdot)$.
These operator-valued distributions are called {\it chronological products} and
verify {\it Bogoliubov axioms}. One starts from a set of {\it interaction
Lagrangians} 
$T_{j}(x), \quad j = 1,\dots, P$ 
and tries to construct the whole series 
$T_{j_{1},\dots,j_{n}}, \quad n \geq 2$.  

Usually, the interactions Lagrangians are Wick monomials.  The canonical
dimension
$\omega(W)$
of certain Wick monomial is defined according to the usual prescription.
By definition, a {\it Wick polynomial} is a sum of Wick monomials.

Bogoliubov axioms are quite natural and they describe the behaviour of the
chronological products with respect to:
(a) the permutation of the couples
$(x_{i},j_{i})$ (symmetry);
(b) the action of the Poincar\'e group in the Fock space of the system
(Poincar\'e invariance);
(c) the factorization property for temporally successive arguments (causality);
(d) the Hermitian conjugation (unitarity).
We mention the essential r\^ole of causality in all approaches to quantum field
theory \cite{Schr1}, \cite{Schr2}.

One considers a {\it interaction Lagrangian} given by a formula of the type:
\be
T(x) = \sum c_{j} T_{j}(x)
\label{one}
\ee
with 
$c_{j}$
some real constants. In this case, the chronological products of the theory are
\be
T(X) = \sum c_{j_{1}} \dots c_{j_{n}} T_{j_{1},\dots,j_{n}}(X)
\label{one-n}
\ee
and they must be plugged into an expression of the type (\ref{S-matrix}) for
$P = 1$
to generate the $S$-matrix of the model.

It can be showed that that one must consider the given interaction 
Lagrangians 
$T_{j}(x)$
to be all Wick monomials canonical dimension 
$\omega_{j} \leq 4$
($j = 1,\dots,P$)
acting in the Fock space of the system.  

If there are non-Hermitian free fields acting in the Fock space, we have in 
general:
\be
T_{j}(x)^{\dagger} = T_{j^{*}}(x)
\label{skew-unitarity1}
\ee
where 
$j \rightarrow j^{*}$
is a bijective map of the numbers
$1,2,\dots,P$.

If there are Fermi or ghost fields acting in the Fock space, the causality
property is in general:
\be
T_{j_{1}}(x_{1}) T_{j_{2}}(x_{2}) = 
(-1)^{\sigma_{j_{1}}\sigma_{j_{2}}} T_{j_{2}}(x_{2}) T_{j_{1}}(x_{1}),
\quad \forall x_{1} \sim x_{2}.
\label{skew-causality1}
\ee

Here
$\sigma_{i}$
is the number of Fermi and ghost fields factors in the Wick monomial
$T_{j}$;
if
$\sigma_{j}$
is even (odd) we call the index
$j$
{\it even} (resp. {\it odd}). One has to keep track of these signs in the
symmetry axiom for the chronological products.

It is convenient to let the index $j$ have the value $0$ also and we
put by definition
\be
T_{0} \equiv {\bf 1}.
\ee

Moreover, we define a new sum operation of two indices 
$j_{1}, j_{2} = 1,\dots,P$;
this summation is denoted by $+$ but should not be confused with the ordinary
sum. By definition we have:
\be
T_{j_{1}+j_{2}}(x) = c :T_{j_{1}}(x) T_{j_{2}}(x):
\ee
for some positive constant $c$. We define componentwise the summation for 
$n$-tuples
$J = \{j_{1},\dots,j_{n}\}$. 
The new summation is non-commutative if Fermi or ghost fields are present.

We will use the notation
\be
\omega_{J} \equiv \sum_{j \in J} \omega_{j}
\ee
and we call it the {\it canonical dimension} of 
$T_{J}(X)$.

According to Epstein-Glaser \cite{EG1} one must add a new axiom, namely the
following {\it Wick expansion} of the chronological products is valid:
\be
T_{J}(X) = \sum_{K+L=J} \epsilon \quad t_{K}(X) \quad W_{L}(X)
\label{wick-t}
\ee
where: (a)
$t_{K}(X)$
are numerical distributions (the {\it renormalized Feynman amplitudes});
(b) the degree of singularity is restricted by the following relation:
\be
\omega(t_{K}) \leq \omega_{K} - 4(n-1);
\label{deg-chrono}
\ee
(c) $\epsilon$ is the sign coming from permutation of Fermi fields;
(d) we have introduced the notation
\be
W_{J}(X) \equiv :T_{j_{1}}(x_{1})\cdots T_{j_{n}}(x_{n}):
\ee

Let us notice that from (\ref{wick-t}) we have:
\be
t_{J}(X) = \left( \Omega, T_{J}(X) \Omega \right)
\label{average-chrono}
\ee 
where $\Omega$ is the vacuum state of the Fock space.

We end this Subsection with an important remark. Let us consider some general
Wick polynomials
\be
A_{i}(x) = \sum_{j} c_{ij} \quad T_{j}(x), \quad i = 1,2,\dots
\ee

Then we can define the chronological products:
\be
T(A_{1}(x_{1}),\cdots,A_{n}(x_{n})) \equiv \sum_{J} 
c_{i_{1}j_{1}} \cdots c_{i_{n}j_{n}} \quad
T_{j_{1},\dots,j_{n}}(x_{1},\cdots, x_{n}).
\label{chrono-arb}
\ee

One can find in \cite{DF1} a system of axioms for the expressions
$T(A_{1}(x_{1}),\cdots,A_{n}(x_{n}))$
which is equivalent to the Bogoliubov set of axioms.

\subsection{Massive Yang-Mills Fields\label{ym}}
         
In \cite{YM} - \cite{fermi} we have justified the following scheme for the
standard model (SM): we consider the auxiliary Hilbert space 
${\cal H}_{YM}^{gh,r}$
generated from the vacuum
$\Omega$
by applying the free fields
$
A_{a\mu},~ u_{a},~\tilde{u}_{a},~\Phi_{a} \quad a = 1,\dots,r
$
where the first one has vector transformation properties with respect to the
Poincar\'e group and the others are scalars. In other words, every vector field
has three scalar partners. Also
$
u_{a},~\tilde{u}_{a} \quad a = 1,\dots,r
$
are Fermion and
$
A_{\mu},~\Phi_{a} \quad a = 1,\dots,r
$
are Boson fields. 

We have two distinct possibilities for distinct indices $a$:

(I) Fields of type I correspond to an index $a$ such that the vector field
$A_{a}^{\mu}$ 
has non-zero mass
$m_{a}$.
In this case we suppose that all the other scalar partners fields
$
u_{a},~\tilde{u}_{a},~\Phi_{a}
$
have the same mass $m_{a}$.

(II) Fields of type II correspond to an index $a$ such that the vector field
$A_{a}^{\mu}$ 
has zero mass. In this case we suppose that the scalar partners fields
$
u_{a},~\tilde{u}_{a}
$
also have the zero mass but the scalar field 
$\Phi_{a}$
can have a non-zero mass: 
$m^{H}_{a} \geq 0$. 
It is convenient to use the compact notation 
\be
m^{*}_{a} \equiv \left\{\begin{array}{rcl} 
m_{a} & \mbox{for} & m_{a} \not= 0 \\
m^{H}_{a} & \mbox{for} & m_{a} = 0\end{array}\right. 
\ee

Then the following following equations of motion describe the preceding
construction: 
\be
(\square + m_{a}^{2}) u_{a}(x) = 0, \quad 
(\square + m_{a}^{2}) \tilde{u}_{a}(x) = 0, \quad
(\square + (m^{*}_{a})^{2}) \Phi_{a}(x) = 0, \quad a = 1,\dots,r.
\label{equ-r}
\ee

We also postulate the following canonical (anti)commutation relations:
\bea
\left[A_{a\mu}(x),A_{b\nu}(y)\right] = - 
\delta_{ab} g_{\mu\nu} D_{m_{a}}(x-y) \times {\bf 1},
\nonumber \\
\{u_{a}(x),\tilde{u}_{b}(y)\} = \delta_{ab} D_{m_{a}}(x-y) \times {\bf 1}, 
\quad
[ \Phi_{a}(x),\Phi_{b}(y) ] = \delta_{ab} D_{m^{*}_{a}}(x-y) \times {\bf 1};
\label{comm-r}
\eea
all other (anti)commutators are null.  
                        
In this Hilbert space we suppose given a sesquilinear form 
$<\cdot, \cdot>$
such that:
\be
A_{a\mu}(x)^{\dagger} = A_{a\mu}(x), \quad
u_{a}(x)^{\dagger} = u_{a}(x), \quad
\tilde{u}_{a}(x)^{\dagger} = - \tilde{u}_{a}(x), \quad
\Phi_{a}(x)^{\dagger} = \Phi_{a}(x).
\label{conjugate-YM}
\ee
                        
The ghost degree is 
$\pm 1$ 
for the fields
$~u_{a}$
(resp.
$
~\tilde{u}_{a}),
\quad a = 1,\dots,r
$
and $0$ for the other fields.

One can define the BRST {\it supercharge} $Q$ by:
\bea
\{Q, u_{a} \}= 0 \quad 
\{Q, \tilde{u}_{a} \}= - i (\partial_{\mu} A_{a}^{\mu} + m_{a} \Phi_{a})
\nonumber \\
~[ Q, A_{a}^{\mu} ] = i \partial^{\mu} u_{a} \quad
[ Q, \Phi_{a} ] = i m_{a} u_{a}, \quad \forall a = 1,\dots,r
\label{BRST-YM}
\eea
and 
\be
Q \Omega = 0.
\ee

Then one can justify that the {\bf physical} Hilbert space of the Yang-Mills
system is a factor space
\be
{\cal H}^{r}_{YM} \equiv {\cal H} \equiv Ker(Q)/Ran(Q).
\ee

The sesquilinear form
$<\cdot, \cdot>$
induces a {\it bona fide} scalar product on the Hilbert factor space.

The factorization process leads to the following {\bf physical} particle
content of this model:
\begin{itemize}
\item
For 
$m_{a} > 0$
the fields
$
A_{a}^{\mu},~u_{a},~\tilde{u}_{a},~\Phi_{a}
$
describe a particle of mass
$m_{a} > 0$
and spin $1$; this are the so-called {\it heavy Bosons} \cite{standard}.
\item
For 
$m_{a} = 0$
the fields
$
A_{a}^{\mu},~u_{a},~\tilde{u}_{a}
$
describe a particle of mass
$0$
and helicity $1$; the typical example is the {\it photon} \cite{YM}.
\item
For 
$m_{a} = 0$
the fields
$
\Phi_{a}
$
describe a scalar fields of mass
$m_{a}^{H}$;
this are the so-called {\it Higgs fields}.
\end{itemize}

This framework is sufficient for the study of the Standard Model (SM) of the
electro-weak interactions. To include also quantum chromodynamics one must
consider that there is a third case:

(III) Fields of type III correspond to an index $a$ such that the vector field
$A_{a}^{\mu}$ 
has zero mass, the scalar partners
$
u_{a},~\tilde{u}_{a}
$
also have zero mass but the scalar field 
$\Phi_{a}$
is absent. 

In \cite{Sc4} and \cite{DS1} the model is constructed somewhat differently: one
eliminates the fields of type II and includes a number of supplementary scalar 
Bosonic fields
$\varphi_{i}$
of masses
$m_{i} \geq 0$.
In this framework one can consider for instance the very interesting
Higgs-Kibble model in which there are no zero-mass particle, so the adiabatic
limit probably exists.

We can preserve the general framework with only two types of indices if we
consider that in case II there are in fact three subcases (i.e three types of
indices $a$ for which 
$m_{a} = 0$):

(IIa) In this case
$A_{a\mu},~ u_{a},~ \tilde{u}_{a},~ \Phi_{a} \not\equiv 0$;

(IIb) In this case
$\Phi_{a} \equiv 0$;

(IIc) In this case
$A_{a\mu},~ u_{a},~ \tilde{u}_{a} \equiv 0$.

One must modify appropriately the canonical (anti)commutation  relations
(\ref{comm-r}) to avoid contradiction for some values of the indices. 
One has some freedom of notation: for instance, one can eliminate case (IIa)
if one includes the first three fields fields in case (IIb) and the last one in
case (IIc). The relations (\ref{BRST-YM}) are not affected in this way.

Let us consider the set of Wick monomials ${\cal W}$ constructed from the free
fields
$A_{a}^{\mu},~u_{a},~\tilde{u}_{a}$
and
$\Phi_{a}$
for all indices
$a = 1,\dots,r$;
we define the BRST operator
$d_{Q}: {\cal W} \rightarrow {\cal W}$
as the (graded) commutator with the supercharge operator $Q$. Then one can
prove easily that:
\be
d_{Q}^{2} = 0.
\label{coomology}
\ee

The class of observables on the factor space is defined as follows: an operator
$O: {\cal H}_{YM}^{gh,r} \rightarrow {\cal H}_{YM}^{gh,r}$
induces a well defined operator 
$[O]$
on the factor space
$\overline{Ker(Q)/Im(Q)} \simeq {\cal F}_{m}$
if and only if it verifies:
\be
\left. d_{Q} O \right|_{Ker(Q)} = 0.
\label{dQ}
\ee
Because of the relation (\ref{coomology}) not all operators verifying the
condition (\ref{dQ}) are interesting. In fact, the operators of the type 
$d_{Q}O$ are inducing a null operator on the factor space; explicitly, we have:
\be
[d_{Q} O] = 0.
\ee

We will construct a perturbation theory verifying Bogoliubov axioms using this
set of free fields and imposing the usual axioms of causality, unitarity and
relativistic invariance on the chronological products
$T(x_{1},\dots,x_{n})$.
Moreover, we want that the result factorizes to the physical Hilbert space in
the formal adiabatic limit. This amounts to \cite{AS} - \cite{DS}:
\be
d_{Q} T(x_{1},\dots,x_{n}) = i \sum_{l=1}^{n} 
{\partial \over \partial x^{\mu}_{l}} T^{\mu}_{l}(x_{1},\dots,x_{n})
\label{gauge-inf}
\ee
for some auxiliary chronological products
$T^{\mu}_{l}(x_{1},\dots,x_{n}), \quad l = 1,\dots,n$
which must be determined recurringly, together with the standard chronological
products.

If one adds matter fields we proceed as before. In particular, we suppose that
the BRST operator acts trivially on the matter fields. It seems that the matter
field must be described by a set of Dirac fields of masses
$M_{A}, \quad A = 1,\dots,N$
denoted by
$\psi_{A}(x)$.
These fields are characterized by the following relations \cite{fermi}; here
$A, B = 1, \dots, N$: 

Equation of motion:
\be
(i \gamma \cdot \partial + M_{A}) \psi_{A}(x) = 0.
\label{dirac-equ-N}
\ee

Canonical (anti)commutation relations: 
\be
\{\psi_{A}(x),\overline{\psi_{B}}(y)\} = \delta_{AB} S_{M_{A}}(x-y) 
\label{CAR-m}
\ee
and all other (anti)commutators are null. 

By a {\it trivial Lagrangian} we mean a Wick expression of the type
\be
L(x) = d_{Q} N(x) + i {\partial \over \partial x^{\mu}} L^{\mu}(x)
\label{T1-trivial}
\ee
with
$L(x)$
and
$L^{\mu}(x)$
some Wick polynomials. The first term in the previous formula gives zero
by factorisation to the physical Hilbert space (according to a previous
discussion) and the second one gives also zero in the adiabatic limit; this
justify the elimination of such expression from the first order chronological
product
$T(x)$.

Let us suppose that for 
$|X| = 1$
the expressions
$T(x)$
and
$T^{\mu}(x) = T^{\mu}_{1}(x)$
have the generic form (\ref{one}):
\be
T(x) = \sum c_{j} T_{j}(x)
\quad
T^{\mu}(x) = \sum c^{\mu}_{j} T_{j}(x)
\label{t1}
\ee
with 
$c_{j}, \quad c^{\mu}_{j}$
some real constants. One can prove \cite{standard}, \cite{fermi} that the 
condition (\ref{gauge-inf}) for
$n = 1, 2, 3$
determines quite drastically the interaction Lagrangian (up to a trivial
Lagrangian):
\bea
T(x) \equiv
- f_{abc} \left[ {1\over 2} :A_{a\mu}(x)A_{b\nu}(x) F_{a}^{\mu\nu}(x):
:A_{a}^{\mu}(x) u_{b}(x) \partial_{\mu} \tilde{u}_{c}(x):\right],
\nonumber \\
+ f'_{abc} \left[ :\Phi_{a}(x) \partial_{\mu} \Phi_{b}(x) A_{c}^{\mu}(x): 
- m_{b} :\Phi_{a}(x) A_{b\mu}(x) A_{c}^{\mu}(x): 
- m_{b} :\Phi_{a}(x) \tilde{u}_{b}(x) u_{c}(x):\right]
\nonumber \\
+ f^{"}_{abc} :\Phi_{a}(x) \Phi_{b}(x) \Phi_{c}(x): 
+  j^{\mu}_{a}(x) A_{a\mu}(x) + j_{a}(x) \Phi_{a}(x)
\label{inter}
\eea
where:
\be
F_{a}^{\mu\nu}(x) \equiv 
\partial^{\mu} A^{\nu}_{a}(x) - \partial^{\nu} A^{\mu}_{a}(x) 
\ee
is the Yang-Mills field tensor and the so-called {\it currents} are:
\be
j_{a}^{\mu}(x) = 
:\overline{\psi_{A}}(x) (t_{a})_{AB} \gamma^{\mu} \psi_{B}(x): +
:\overline{\psi_{A}}(x) (t'_{a})_{AB} \gamma^{\mu} \gamma_{5} \psi_{B}(x):
\label{vector-current}
\ee
and
\be
j_{a}(x) = 
:\overline{\psi_{A}}(x) (s_{a})_{AB} \psi_{B}(x): +
:\overline{\psi_{A}}(x) (s'_{a})_{AB} \gamma_{5} \psi_{B}(x):
\label{scalar-current}
\ee
where a number of restrictions must be imposed on the various constants (see
\cite{YM}-\cite{fermi} where the condition of gauge invariance is analysed up
to order $3$.)

Moreover, we can take 
$T^{\mu}(x)$
to be:
\bea
T^{\mu}(x)  = f_{abc} \left[ :u_{a}(x) A_{b\nu}(x) F^{\nu\mu}_{c}(x): -
{1\over 2} :u_{a}(x) u_{b}(x) \partial^{\mu}(x) \tilde{u}_{c}(x): \right]
\nonumber \\
+ f'_{abc} \left[ m_{a} :A_{a}^{\mu}(x) \Phi_{b}(x) u_{c}(x):
+ :\Phi_{a}(x) \partial^{\mu}\Phi_{b}(x) u_{c}(x): \right].
+  u_{a}(x) j^{\mu}_{a}(x).
\label{inter-mu}
\eea

The expressions
$T(x)$
and
$T^{\mu}(x)$
are
$SL(2,\C)$-covariant, are causally commuting and are Hermitean. Moreover we
have the following ghost content:
\be
gh(T(x)) = 0, \quad gh(T^{\mu}(x)) = 0.
\ee

\begin{rem}
The presence of indices of type IIb and IIc is taken into account by requiring
that the constants from 
$T(x)$
are null if one of the indices 
$a, b, c$
takes such values. One can see that this does not affect the equations from the
statement of the theorem.
\end{rem}

\newpage
\section{The Renormalizability of Quantum Electrodynamics}
\subsection{The General Setting}

The case of QED is a particular case of the scheme described in the preceding
Section. We have only one field of type IIb i.e. the triplet
$A_{\mu},~u,~\tilde{u}$
of null mass; they describe a system of null-mass Bosons of helicity $1$ (i.e.
{\it photons}). We also have only one Dirac field
$\psi$
describing the electron. We suppose that in the Hilbert space
${\cal H}^{gh}$
generated by these fields from the vacuum
$\Omega$
we have a sesqui-linear form
$<\cdot,\cdot>$
and we denote the conjugate of the operator $O$ with respect to this form by 
$O^{\dagger}$.
We characterize this form by requiring:
\be
A_{\mu}(x)^{\dagger} = A_{\mu}(x), \quad
u(x)^{\dagger} = u(x), \quad
\tilde{u}(x)^{\dagger} = - \tilde{u}(x).
\label{conjugate}
\ee

The unitary operator realizing the charge conjugation is defined by:
\bea
U_{C} A^{\mu}(x) U_{C}^{-1} = - A^{\mu}(x),
\quad
U_{C} u(x) U_{C}^{-1} = - u(x),
\quad
U_{C} \tilde{u}(x) U_{C}^{-1} = - \tilde{u}(x),
\nonumber \\
U_{C} \psi(x) U_{C}^{-1} = \gamma_{0} \gamma_{2} \bar{\psi}(x)^{t}, \quad
U_{C} \Omega = \Omega
\label{charge}
\eea

Now, we define in
${\cal H}^{gh}$
the {\it supercharge} according to:
\be
Q \Omega = 0
\label{Q-0}
\ee
and
\be
\{Q,u(x)\} = 0,\quad
\{Q,\tilde{u}(x)\} =  - i \partial^{\mu} A_{\mu}(x),\quad
[Q, A_{\mu}(x)] = i \partial_{\mu} u(x).
\label{Q-com}
\ee

The expression of the BRST-operator
$d_{Q}$
follows as a particular case of the corresponding formul\ae~of the Yang-Mills
case. From these properties one can derive
\be
Q^{2} = 0;
\label{square}
\ee
so we also have
\be
Im(Q) \subset Ker(Q).
\label{im-ker}
\ee

By definition, the interaction Lagrangian is:
\be
T(x) \equiv e :\bar{\psi}(x) \gamma_{\mu} \psi(x): A^{\mu}(x)
\label{T1}
\ee
(here $e$ is a real constant: the electron charge) and one can verify easily
that we have the covariance properties with respect to
$SL(2,\C)$.
The most important property is (\ref{gauge-inf}) for 
$n = 1$:
\be
d_{Q} T(x) = i {\partial\over \partial x^{\mu}} T^{\mu}(x)
\label{gauge-inf-1}
\ee
with:
\be
T^{\mu}(x) \equiv e :\bar{\psi}(x) \gamma^{\mu} \psi(x): u(x).
\label{T1/1}
\ee

One can easily check that we have charge-conjugation invariance in the sense:
\be
U_{C} T(x) U^{-1}_{C} = T(x), \quad
U_{C} T^{\mu}(x) U^{-1}_{C} = T^{\mu}(x).
\label{charge-inv}
\ee

We note that we also have:
\be
d_{Q} T^{\mu}(x) = 0.
\label{gauge-prime}
\ee

\subsection{The Main Result}

It is convenient to write the formul\ae~ (\ref{gauge-inf-1}) and
(\ref{gauge-prime}) in a compact way as follows. One denotes by
$A^{k}(x),~ k = 1,\dots,5$
the expressions
$T(x),~T^{\mu}(x)$;
that is, the index $i$ can take the values
$L, \quad \mu$
according to the identification:
$
A^{L}(x) \equiv T(x), \quad A^{\mu}(x) \equiv T^{\mu}(x).
$
Then we can write the preceding gauge invariance conditions in the form:
\be
d_{Q} A^{k}(x) = i \sum_{m=1}^{5} c^{k;\mu}_{m}
{\partial\over \partial x^{\mu}} A^{m}(x), \quad k = 1,\dots,5
\label{gauge-inf-comp-1}
\ee
for some constants
$c^{k;\mu}_{m}$;
the explicit expressions can be obtained from the corresponding gauge
conditions. Only the expression
\be
c^{L;\mu}_{\nu} \equiv \delta^{\mu}_{\nu}
\label{const}
\ee
are non-zero. Then we can prove the following result:
\begin{thm}
One can chose the chronological products such that, beside the fulfilment of
the Bogoliubov axioms, the following identities are verified:
\be
d_{Q} T(A^{k_{1}}(x_{1}),\dots,A^{k_{p}}(x_{p})) = 
i \sum_{l=1}^{p} (-1)^{s_{l}} \sum_{m} c^{k_{l};\mu}_{m}
{\partial\over \partial x^{\mu}_{l}} 
T(A^{k_{1}}(x_{1}),\dots,A^{m}(x_{l}),\dots,A^{k_{p}}(x_{p}))
\label{gauge-inf-comp}
\ee
for all 
$p \in \N$
and all
$k_{1}, \dots, k_{p} = 1,\dots,5$.
Here we have denoted
\be
s_{0} \equiv 0, \quad s_{l} \equiv \sum_{j=1}^{l-1} gh(A_{j}), \quad 
\forall l = 1,\dots,p. 
\ee
\end{thm}

{\bf Proof:} (i) We use induction. Suppose we have constructed the
chronological products such that that all conditions are satisfied up to order
$p = n - 1$.
One can construct the chronological products in order $n$ such that all
Bogoliubov axioms are satisfied, except the condition of gauge invariance.
This can be done directly from the Epstein-Glaser methods \cite{qed} or using
the extension method \cite{DF1}. One can choose the chronological products to
depend only on the fields
\be
A_{\mu},~u,~\psi,~\bar{\psi}
\label{C}
\ee
and such that
\be
gh(T(A^{k_{1}}(x_{1}),\dots,A^{k_{p}}(x_{p}))) = 
\sum_{l=1}^{n} gh(A^{k_{l}}).
\label{Gh}
\ee

Moreover, the symmetry axiom implies relations of the type:
\be
T(A_{1}(x_{1}),A_{2}(x_{2}),\dots) = (-1)^{gh(A_{1}) gh(A_{2})}
T(A_{2}(x_{2}),A_{1}(x_{1}),\dots).
\label{S}
\ee

From the induction procedure (or the extension method) one can easily prove the
the possible obstructions to the gauge invariance condition
(\ref{gauge-inf-comp}) in order $n$ have a particular structure. We have:
\bea
d_{Q} T(A^{k_{1}}(x_{1}),\dots,A^{k_{n}}(x_{n}))
\nonumber \\
= i \sum_{l=1}^{n} (-1)^{s_{l}} \sum_{m} c^{k_{l};\mu}_{m}
{\partial\over \partial x^{\mu}_{l}} 
T(A^{k_{1}}(x_{1}),\dots,A^{m}(x_{l}),\dots,A^{k_{n}}(x_{n}))
\nonumber \\
+ P^{k_{1},\dots,k_{n}}(x_{1},\dots,x_{n})
\label{gauge-inf-comp-n}
\eea
where
$P^{\dots}(X) \equiv P^{\dots}(x_{1},\dots,x_{n})$
are quasi-local operators called {\it anomalies}. They have the following
structure:
\be
P(X) = \sum_{L} \left[ p_{L}(\partial) \delta(X)\right] W_{L}(X)
\label{wickP}
\ee
where
$W_{L}$
are Wick monomials and
$p_{L}$
are polynomials in the derivatives of the type
\be
p_{L}(X) = \sum_{|\alpha| \leq deg(p_{L})} c_{L,\alpha} \partial^{\alpha}
\ee
with the maximal degree restricted by
\be
deg(p_{L}) + \omega_{L} \leq 5.
\label{degP}
\ee

Moreover, we can easily obtain:
\be
gh(P^{k_{1},\dots,k_{n}}(X)) = \sum_{l=1}^{n} gh(A^{k_{l}}) + 1.
\label{Gh'}
\ee

Finally, the anomalies can be chosen
$SL(2,\C)$-covariant and charge conjugation invariant:
\be
U_{C} P^{k_{1},\dots,k_{n}}(X) U_{C}^{-1} = P^{k_{1},\dots,k_{n}}(X).
\label{charge-ano}
\ee

(ii) We have a lot of restrictions on the anomalies. The most sever one comes
from from (\ref{degP}) and (\ref{Gh'}): we obtain that for
\be
\sum_{l=1}^{n} gh(A^{k_{l}}) \geq 5
\label{Gh"}
\ee
there are no anomalies. From this restriction it follows that we have the
following set of relations with possible anomalies:
\bea
d_{Q} T(T(x_{1}),\dots,T(x_{n}))
\nonumber \\
= i \sum_{l=1}^{n} {\partial\over \partial x^{\mu}_{l}} 
T(T(x_{1}),\dots,T^{\mu}(x_{l}),\dots,T(x_{n}))
+ P_{1}(x_{1},\dots,x_{n})
\label{qed1}
\eea
\bea
d_{Q} T(T^{\mu}(x_{1}),T(x_{2}),\dots,T(x_{n}))
\nonumber \\
= -i \sum_{l=2}^{n} {\partial\over \partial x^{\nu}_{l}} 
T(T^{\mu}(x_{1}),T(x_{2}),\dots,T^{\nu}(x_{l}),\dots,T(x_{n}))
\nonumber \\
+ P^{\mu}_{2}(x_{1},\dots,x_{n})
\label{qed2}
\eea
\bea
d_{Q} T(T^{\mu}(x_{1}),T^{\nu}(x_{2}),T(x_{3}),\dots,T(x_{n}))
\nonumber \\
= i \sum_{l=3}^{n} {\partial\over \partial x^{\rho}_{l}} 
T(T^{\mu}(x_{1}),T^{\nu}(x_{2}),T(x_{3}),\dots,T^{\rho}(x_{l}),\dots,
T(x_{n})) 
\nonumber \\
+ P^{\mu\nu}_{3}(x_{1},\dots,x_{n})
\label{qed3}
\eea
\bea
d_{Q} T(T^{\mu}(x_{1}),T^{\nu}(x_{2}),T^{\rho}(x_{3}),T(x_{4}),\dots,T(x_{n}))
\nonumber \\
= - i \sum_{l=4}^{n} {\partial\over \partial x^{\sigma}_{l}} 
T(T^{\mu}(x_{1}),T^{\nu}(x_{2}),T^{\rho}(x_{3}),T(x_{4}),\dots,
T^{\sigma}(x_{l}),\dots,T(x_{n})) 
\nonumber \\
+ P^{\mu\nu\rho}_{4}(x_{1},\dots,x_{n})
\label{qed4}
\eea
\bea
d_{Q} T(T^{\mu}(x_{1}),T^{\nu}(x_{2}),T^{\rho}(x_{3}),T^{\sigma}(x_{4}),\dots,
T(x_{n}))
\nonumber \\
= i \sum_{l=5}^{n} {\partial\over \partial x^{\lambda}_{l}} 
T(T^{\mu}(x_{1}),T^{\nu}(x_{2}),T^{\rho}(x_{3}),T^{\sigma}(x_{4}),\dots,
T^{\lambda}(x_{l}),\dots,T(x_{n})) 
\nonumber \\
+ P^{\mu\nu\rho\lambda}_{5}(x_{1},\dots,x_{n})
\label{qed5}
\eea
where we use, as before, the convention
$\sum_{\emptyset} \equiv 0$.
We can assume that:
\be
P^{\mu\nu}_{3}(X) = 0, \quad |X| = 1, \quad
P^{\mu\nu\rho}_{4}(X) = 0, \quad |X| = 2, \quad
P^{\mu\nu\rho}_{5}(X) = 0, \quad |X| = 3
\ee
without losing generality. The anomalies verify the restrictions (\ref{Gh'})
and (\ref{degP}) and they depend only on the fields
\be
A_{\mu},~\partial_{\nu}A_{\mu},~u,~\partial_{\nu}u,~\psi,~\partial_{\nu}\psi,
\bar{\psi},~\partial_{\nu}\bar{\psi}
\label{C'}
\ee

In fact, we can refine the induction hypothesis: we can assume that
$T(T(x_{1}),\dots,T(x_{n}))$
does not depend on $u$,
$T(T^{\mu}(x_{1}),T(x_{2}),\dots,T(x_{n}))$
depends linearly on
$u(x_{1})$
and does not depend on
$A_{\mu}(x_{1})$,
$T(T^{\mu}(x_{1}),T^{\nu}(x_{2}),T(x_{3}),\dots,T(x_{n}))$
depends linearly on
$:u(x_{1}) u(x_{2}):$
and does not depend on
$A_{\mu}(x_{i}), \quad i = 1,2$,
etc.

Then it follows that the anomalies depend on the fields
\be
A_{\mu},~u,~\partial_{\nu}u,~\psi,~\partial_{\nu}\psi,
\bar{\psi},~\partial_{\nu}\bar{\psi}.
\label{C"}
\ee

From (\ref{S}), we get the following symmetry properties:
\be
P_{1}(x_{1},\dots,x_{n}) \quad
{\rm is~ symmetric~ in} \quad x_{1},\dots,x_{n};
\label{S1}
\ee
\be
P_{2}^{\mu}(x_{1},\dots,x_{n}) \quad
{\rm is~ symmetric~ in} \quad x_{2},\dots,x_{n};
\label{S2}
\ee
\be
P_{3}^{\mu\nu}(x_{1},\dots,x_{n}) \quad
{\rm is~ symmetric~ in} \quad x_{3},\dots,x_{n};
\label{S3}
\ee
\be
P_{4}^{\mu\nu\rho}(x_{1},\dots,x_{n}) \quad
{\rm is~ symmetric~ in} \quad x_{4},\dots,x_{n};
\label{S4}
\ee
\be
P_{5}^{\mu\nu\rho\sigma}(x_{1},\dots,x_{n}) \quad
{\rm is~ symmetric~ in} \quad x_{5},\dots,x_{n};
\label{S5}
\ee
\be
P_{3}^{\mu\nu}(x_{1},\dots,x_{n}) \quad
{\rm is~ antisymmetric~ in} \quad (x_{1},\mu), (x_{2},\nu);
\label{S3'}
\ee
\be
P_{4}^{\mu\nu\rho}(x_{1},\dots,x_{n}) \quad
{\rm is~ antisymmetric~ in} \quad (x_{1},\mu), (x_{2},\nu), (x_{3},\rho);
\label{S4'}
\ee
\be
P_{5}^{\mu\nu\rho\sigma}(x_{1},\dots,x_{n}) \quad
{\rm is~ antisymmetric~ in} \quad (x_{1},\mu), (x_{2},\nu), (x_{3},\rho),
(x_{4},\sigma).
\label{S5'}
\ee

(iii) If we apply the operator
$d_{Q}$
to the anomalous relations (\ref{qed1})-(\ref{qed5}) we easily obtain some
consistency relations quite analogous to the well-known Wess-Zumino consistency
relations:
\be
d_{Q} P_{1}(x_{1},\dots,(x_{n}) 
= i \sum_{l=1}^{n} {\partial\over \partial x^{\mu}_{l}} 
P^{\mu}_{2}(x_{l},x_{1},\dots,\hat{x}_{l},\dots,x_{n})
\label{qedG1}
\ee
\be
d_{Q} P^{\mu}_{2}(x_{1}),\dots,(x_{n})
= -i \sum_{l=2}^{n} {\partial\over \partial x^{\nu}_{l}} 
P^{\mu\nu}_{3}(x_{1},x_{l},x_{2},\dots,\hat{x}_{l},\dots,x_{n})
\label{qedG2}
\ee
\be
d_{Q} P^{\mu\nu}_{3}(x_{1},\dots,x_{n})
= i \sum_{l=3}^{n} {\partial\over \partial x^{\rho}_{l}}
P^{\mu\nu\rho}_{4}(x_{1},x_{2},x_{l},x_{3},\dots,\hat{x}_{l},\dots,x_{n})
\label{qedG3}
\ee
\be
d_{Q} P^{\mu\nu\rho}_{4}(x_{1},x_{n})
= - i \sum_{l=4}^{n} {\partial\over \partial x^{\sigma}_{l}} 
P^{\mu\nu\rho\sigma}_{5}(x_{1},x_{2},x_{3},x_{l},x_{4},\dots,\hat{x}_{l},
\dots,x_{n})
\label{qedG4}
\ee
\be
d_{Q} P^{\mu\nu\rho\sigma}_{5}(x_{1},\dots,x_{n}) = 0.
\label{qedG5}
\ee

We will use repeatedly the identity
\be
\sum_{i=1}^{n} {\partial\over \partial x^{\rho}_{l}} \delta(X) = 0
\label{delta}
\ee
where
\be
\delta(X) \equiv \delta(x_{1}-x_{n}) \cdots \delta(x_{n-1}-x_{n}).
\ee

If we take into account (\ref{degP}) and (\ref{Gh'}), the generic form of the
anomalies is:
\bea
P_{l}^{\dots}(X) = \delta(X) \widetilde{W}_{l}^{\dots}(x_{1})
+ \sum_{p=1}^{n} \left[ {\partial\over \partial x^{\mu}_{p}} \delta(X)\right]
\widetilde{W}_{l;p}^{\dots;\mu}(X)
+ \sum_{p,q=1}^{n} \left[ 
{\partial^{2}\over \partial x^{\mu}_{p} \partial x^{\nu}_{q}} \delta(X)\right]
\widetilde{W}_{l;pq}^{\dots;\mu\nu}(X)
\nonumber\\
+ \sum_{p,q,r=1}^{n} \left[ 
{\partial^{2}\over \partial x^{\mu}_{p} \partial x^{\nu}_{q} 
\partial x^{\rho}_{r}} \delta(X)\right] 
\widetilde{W}_{l;pqr}^{\dots;\mu\nu\rho}(X)
+ \sum_{p,q,r,s=1}^{n} \left[ 
{\partial^{2}\over \partial x^{\mu}_{p} \partial x^{\nu}_{q} 
\partial x^{\rho}_{r} \partial x^{\sigma}_{s}} \delta(X)\right] 
\widetilde{W}_{l;pqrs}^{\dots;\mu\nu\rho\sigma}(X) \qquad
\eea
where
$\widetilde{W}^{\dots}_{\dots}$
are some Wick polynomials with convenient symmetry properties. If we use
(\ref{delta}) we can eliminate all derivatives with respect to one variable,
say
$x_{1}$
if we redefine conveniently the expressions
$\widetilde{W}^{\dots}_{\dots}$:
\bea
P_{l}^{\dots}(X) = \delta(X) W_{l}^{\dots}(x_{1})
+ \sum_{p=2}^{n}  {\partial\over \partial x^{\mu}_{p}} \delta(X)
W_{l;p}^{\dots;\mu}(x_{1})
+ \sum_{p,q=2}^{n} 
{\partial^{2}\over \partial x^{\mu}_{p} \partial x^{\nu}_{q}} \delta(X)
W_{l;pq}^{\dots;\mu\nu}(x_{1})
\nonumber\\
+ \sum_{p,q,r=2}^{n}
{\partial^{2}\over \partial x^{\mu}_{p} \partial x^{\nu}_{q} 
\partial x^{\rho}_{r}} \delta(X) W_{l;pqr}^{\dots;\mu\nu\rho}(x_{1})
+ \sum_{p,q,r,s=2}^{n}
{\partial^{2}\over \partial x^{\mu}_{p} \partial x^{\nu}_{q} 
\partial x^{\rho}_{r} \partial x^{\sigma}_{s}} \delta(X)
W_{l;pqrs}^{\dots;\mu\nu\rho\sigma}(x_{1})
\label{A}
\eea

If
$f \in {\cal S}(\R^{4n})$
is arbitrary we have:
\bea
<P^{\dots}_{l},f(X)> = \int dx f(x,\dots,x) W^{\dots}_{l}(x)
- \sum_{p=2}^{n} \int dx (\partial^{p}_{\mu}f)(x,\dots,x) 
W^{\dots;\mu}_{l;p}(x)
\nonumber \\
+ \sum_{p,q=2}^{n} \int dx 
(\partial^{p}_{\mu}\partial^{q}_{\nu}f)(x,\dots,x) W^{\dots;\mu\nu}_{l;pq}(x)
+ \cdots
\eea

But the expressions
$
f(x,\dots,x),~(\partial^{p}_{\mu}f)(x,\dots,x),~
(\partial^{p}_{\mu}\partial^{q}_{\nu}f)(x,\dots,x),\dots \quad
p, q, \dots \geq 2
$
can be chosen arbitrary, so we have:
\be
P^{\dots}_{l}(X) \quad \Longleftrightarrow \quad W^{\dots}_{l} = 0, \quad
W^{\dots;\mu}_{l;p}(X) = 0, \quad W^{\dots;\mu\nu}_{l;pq}(X) = 0, \quad, \dots
\ee

As a consequence, every symmetry property
\be
< P^{\dots}_{l}(X), f^{g}(X) > = < P^{\dots}_{l}(X), f(X) >
\ee
for $g$ an arbitrary symmetry, will be equivalent to corresponding symmetry
properties for the Wick polynomials:
\be
g\cdot W = W.
\label{S'}
\ee

(iv) Let us consider 
$l = 3, 4, 5$;
because in this case
$
gh(P_{l}^{\dots}) \geq 3
$
in every Wick polynomial
$W^{\dots}_{\dots}$
from (\ref{A}) we have at least two factors $u$ (the third can be
$\partial u$)
so we get:
\be
P_{l}^{\dots}(X) = 0, \quad l = 3,4,5.
\ee

The generic expression of
$P_{2}$
is:
\bea
P_{2}^{\mu}(X) = \delta(X) W_{2}^{\mu}(x_{1})
+ \sum_{p=2}^{n}  {\partial\over \partial x^{\nu}_{p}} \delta(X)
W_{2;p}^{\mu;\nu}(x_{1})
+ \sum_{p,q=2}^{n} 
{\partial^{2}\over \partial x^{\nu}_{p} \partial x^{\rho}_{q}} \delta(X)
W_{2;pq}^{\mu;\nu\rho}(x_{1})
\nonumber\\
+ \sum_{p,q,r=2}^{n}
{\partial^{3}\over \partial x^{\nu}_{p} \partial x^{\rho}_{q} 
\partial x^{\sigma}_{r}} \delta(X) W_{2;pqr}^{\mu;\nu\rho\sigma}(x_{1})
\eea

Because
$gh(P_{2}) = 2$
we have
$W_{l;pqr}^{\mu;\nu\rho\sigma} \sim :uu: = 0$
so the last term disappears. If we use the symmetry property (\ref{S2}) we get:
\bea
W_{2;p}^{\mu;\nu} = W_{2;2}^{\mu;\nu} \equiv W_{2}^{\mu;\nu},
\quad \forall p = 2,\dots,n,
\nonumber \\
W_{2;pq}^{\mu;\nu\rho} = W_{2;22}^{\mu;\nu\rho} \equiv W_{2}^{\mu;\nu\rho}, 
\quad \forall p,q = 2,\dots,n
\eea
so we can write the preceding expression more simply:
\bea
P_{2}^{\mu}(X) = \delta(X) W_{2}^{\mu}(x_{1})
+ \sum_{p=2}^{n}  {\partial\over \partial x^{\nu}_{p}} \delta(X)
W_{2}^{\mu;\nu}(x_{1})
+ \sum_{p,q=2}^{n} 
{\partial^{2}\over \partial x^{\nu}_{p} \partial x^{\rho}_{q}} \delta(X)
W_{2}^{\mu;\nu\rho}(x_{1}).
\eea

If we use (\ref{delta}) we obtain after some relabelling:
\bea
P_{2}^{\mu}(X) = \delta(X) W_{2}^{\mu}(x_{1})
+ {\partial\over \partial x^{\nu}_{1}} 
\left[ \delta(X) W_{2}^{\mu;\nu}(x_{1})\right]
+ {\partial^{2}\over \partial x^{\nu}_{1} \partial x^{\rho}_{1}} 
\left[ \delta(X) W_{2}^{\mu;\nu\rho}(x_{1}) \right]
\label{P2'}
\eea
and we can assume that
\be
W_{2}^{\mu;\nu\rho} = (\nu \leftrightarrow \rho).
\ee

Because
$gh(P_{1}) = 1$
we have the generic expression:
\bea
P_{1}(X) = \delta(X) W_{1}(x_{1})
+ \sum_{p=2}^{n}  {\partial\over \partial x^{\mu}_{p}} \delta(X)
W_{1;p}^{\mu}(x_{1})
+ \sum_{p,q=2}^{n} 
{\partial^{2}\over \partial x^{\mu}_{p} \partial x^{\nu}_{q}} \delta(X)
W_{1;pq}^{\mu\nu}(x_{1})
\nonumber\\
+ \sum_{p,q,r=2}^{n}
{\partial^{3}\over \partial x^{\mu}_{p} \partial x^{\nu}_{q} 
\partial x^{\rho}_{r}} \delta(X) W_{1;pqr}^{\mu\nu\rho}(x_{1})
+ \sum_{p,q,r,s=2}^{n}
{\partial^{4}\over \partial x^{\mu}_{p} \partial x^{\nu}_{q} 
\partial x^{\rho}_{r} \partial x^{\sigma}_{s}} \delta(X)
W_{1;pqrs}^{\mu\nu\rho\sigma}(x_{1})
\label{P1}
\eea

The symmetry requirement (\ref{S1}) in
$x_{2},\dots,x_{n}$
leads as above at a simpler form:
\bea
P_{1}(X) = \delta(X) W_{1}(x_{1})
+ {\partial\over \partial x^{\mu}_{1}} 
\left[ \delta(X) W_{1}^{\mu}(x_{1}) \right]
+ {\partial^{2}\over \partial x^{\mu}_{1} \partial x^{\nu}_{1}} 
\left[ \delta(X) W_{1}^{\mu\nu}(x_{1}) \right]
\nonumber\\
+ {\partial^{3}\over \partial x^{\mu}_{1} \partial x^{\nu}_{1} 
\partial x^{\rho}_{1}} \left[ \delta(X) W_{1}^{\mu\nu\rho}(x_{1})\right]
+ {\partial^{4}\over \partial x^{\mu}_{1} \partial x^{\nu}_{1} 
\partial x^{\rho}_{1} \partial x^{\sigma}_{1}} 
\left[ \delta(X) W_{1}^{\mu\nu\rho\sigma}(x_{1})\right]
\label{P1'}
\eea
and the Wick polynomials have convenient symmetry properties. We have the
generic form
\be
W_{1}^{\mu\nu\rho\sigma} = c_{1}^{\mu\nu\rho\sigma} u
\ee
with
$c_{1}^{\mu\nu\rho\sigma}$
a Lorentz covariant tensor. If we perform the finite renormalization:
\be
T(T^{\mu}(x_{1}),T(x_{2}),\dots,T(x_{n})) \rightarrow
T(T^{\mu}(x_{1}),T(x_{2}),\dots,T(x_{n})) 
+ i {\partial^{3}\over \partial x^{\nu}_{1} \partial x^{\rho}_{1} 
\partial x^{\sigma}_{1}} \left[ \delta(X) W_{1}^{\mu\nu\rho\sigma}(x_{1})
\right] 
\label{R1}
\ee
we do not affect the symmetry properties and the field dependence (\ref{C"})
but as a result we eliminate the last term in the expression of
$P_{1}$.
We impose now the symmetry property (\ref{S1}) in
$x_{1}, x_{2}$
and obtain that in fact:
\be
P_{1}(X) = \delta(X) W_{1}(x_{1}).
\label{P1"}
\ee

(v) Next, we use the consistency conditions (\ref{qedG1})-(\ref{qedG5}). Only
the first two one are non-trivial. We get the following conditions:
\be
d_{Q} W_{1} = i \partial_{\mu} W^{\mu}_{2}, \quad
W^{\mu;\nu}_{2} = - (\mu \leftrightarrow \nu), \quad
W^{\mu;\nu\rho}_{2} + W^{\nu;\mu\rho}_{2} + W^{\rho;\mu\nu}_{2} = 0
\label{G1"}
\ee
and respectively:
\be
d_{Q} W^{\mu}_{2} = 0, \quad
d_{Q} W^{\mu;\nu}_{2} = 0, \quad
d_{Q} W^{\mu;\nu\rho}_{2} = 0.
\label{G2"}
\ee

We can still simplify the expressions of the anomalies by finite
renormalizations. We present briefly the details. The generic form of
$W^{\mu;\nu\rho}_{2}$
is
\be
W^{\mu;\nu\rho}_{2} = c^{\mu\nu\rho\sigma}_{2} :u \partial_{\sigma}u:
\ee
with
$c^{\mu\nu\rho\sigma}_{2}$
a Lorentz invariant tensor. If we define
\be
U^{\mu;\nu\rho}_{2} = c^{\mu\nu\rho\sigma}_{2} : u A_{\sigma}:
\ee
then we have:
\be
d_{Q} U^{\mu;\nu\rho}_{2} = - i W^{\mu;\nu\rho}_{2}.
\ee

It follows that if we perform the finite renormalization:
\be
T(T^{\mu}(x_{1}),T(x_{2}),\dots,T(x_{n})) \rightarrow
T(T^{\mu}(x_{1}),T(x_{2}),\dots,T(x_{n})) 
+ i {\partial^{2}\over \partial x^{\nu}_{1} \partial x^{\rho}_{1}}
\left[ \delta(X) U_{2}^{\mu\nu\rho}(x_{1}) \right] 
\label{R2}
\ee
we do not change the symmetry properties and the field structure; moreover we
do not enter in conflict with (\ref{R1}). As a result we make
\be
W^{\mu;\nu\rho}_{2} = 0.
\ee

In the same way, we have the generic expression:
\be
W^{\mu;\nu}_{2} = \tilde{c}^{\mu\nu\rho\sigma}_{2} 
:u \partial_{\rho}u A_{\sigma}:
\ee
with
$\tilde{c}^{\mu\nu\rho\sigma}_{2}$
a Lorentz invariant tensor. From the second equation (\ref{G1"}) we obtain
antisymmetry in the first two indices and from the second equation (\ref{G2"})
we get symmetry in the last two indices. All these restrictions lead to
\be
\tilde{c}^{\mu\nu\rho\sigma}_{2} = 0.
\ee

So, in the end we have:
\be
P^{\mu}_{2} = \delta(X) W^{\mu}_{2}(x_{1}).
\ee

From the equations (\ref{G1"}) and (\ref{G2"}) we are left with:
\be
d_{Q} W_{1} = i \partial_{\mu} W^{\mu}_{2}, \quad
d_{Q} W^{\mu}_{2} = 0.
\label{G'}
\ee

(vi) We have now the generic form:
\be
W^{\mu}_{1} = d_{1} :u \partial^{\mu}u: 
+ d_{2} :u \partial^{\mu}u A_{\rho} A^{\rho}:
+ d_{3} :u \partial_{\rho}u A_{\rho} A^{\mu}:
\ee
for some constants
$d_{i}$.
The second equation (\ref{G'}) gives
$d_{3} = 2 d_{1}$.
If we define:
\be
U^{\mu}_{2} = d_{1} :u A^{\mu}: 
+ d_{2} :u A^{\mu}u A_{\rho} A^{\rho}:
\ee
we get
\be
d_{Q} U^{\mu}_{2} = i W^{\mu}_{2}.
\ee
Now we perform the finite renormalization
\be
T(T^{\mu}(x_{1}),T(x_{2}),\dots,T(x_{n})) \rightarrow
T(T^{\mu}(x_{1}),T(x_{2}),\dots,T(x_{n})) 
+ i \delta(X) U_{2}^{\mu}(x_{1})
\label{R3}
\ee
we do not affect the properties of the chronological products, we do not spoil
the previous two renormalizations and we make:
\be
P^{\mu}_{2} = 0.
\ee

It follows that we still have to impose:
\be
d_{Q} W_{1} = 0.
\label{G"}
\ee

The generic form of
$W_{1}$
is:
\bea
W_{1} = c_{1} u + c_{2} :u A_{\mu} A^{\mu}: + c_{3} :\partial_{\mu}u A^{\mu}:
+ c_{4} :u \bar{\psi} \psi: + c_{5} :u \bar{\psi} \gamma_{5} \psi:
\nonumber \\
+ c_{6} :u A_{\mu} \bar{\psi} \gamma^{\mu} \psi: 
+ c_{7} :u A_{\mu} \bar{\psi} \gamma^{\mu} \gamma_{5} \psi:
+ c_{8} :\partial_{\mu}u A^{\mu} A_{\rho} A^{\rho}:
+ c_{9} :u A^{\mu} A^{\mu} A_{\rho} A^{\rho}:
\label{W1}
\eea

Now it is time to use charge conjugation invariance of the anomalies
(\ref{charge-ano}) for
$P_{1}$;
we get easily:
$c_{i} = 0, \quad i = 1, 2, 4, 5, 7, 9$.
If we impose the condition (\ref{G"}) we get
$c_{6} = 0$.
It follows that we are left with:
\be
W_{1} = c_{3} :\partial_{\mu}u A^{\mu}:
+ c_{9} :\partial_{\mu}u A^{\mu} A_{\rho} A^{\rho}:
\label{W1'}
\ee

If we define:
\be
U_{1} \equiv {1\over 2} c_{3} :A_{\mu} A^{\mu}:
+ {1\over 4} c_{9} :A_{\mu} A^{\mu} A_{\rho} A^{\rho}:
\ee
we have:
\be
d_{Q} U_{1} = i W_{1}
\ee

Finally, we preform the finite renormalization:
\be
T(T(x_{1}),\dots,T(x_{n})) \rightarrow T(T(x_{1}),\dots,T(x_{n})) 
+ i \delta(X) U_{1}(x_{1})
\label{R4}
\ee
we do not affect the symmetry properties and the field structure (\ref{C"}). As
a result we get:
\be
P_{1}(X) = 0
\ee
and the proof is finished.
$\qed$
\begin{rem}
It is easy to see that the same pattern works for scalar electrodynamics also.
A minor modification appears for the expression of
$W_{1}$:
the terms
$c_{4}-c_{7}$
must be replaced by:
\be
W_{1} = c_{4} :u \bar{\phi} \phi:
+ c_{5} :u A_{\mu} \bar{\phi} \partial^{\mu} \phi:
+ c_{6} :u A_{\mu} \partial^{\mu}\bar{\phi} \phi:
\ee
The first contribution is cancelled by charge conjugation invariance and the
last two by the condition (\ref{G"}).
\end{rem}

\newpage
\section{The Structure of the Anomalies in Higher Orders\label{ano}}

\subsection{The Anomalous Gauge Equations}

We give now the results for the Yang-Mills model as presented in Subsection 
\ref{ym}. By comparison to the case of QED, two important modification appear.
The first one is the relation (\ref{gauge-prime}) which is replaced by:
\be
d_{Q} T^{\mu}(x) = i {\partial \over \partial x^{\nu}} T^{\mu\nu}(x)
\label{gauge-prime-ym}
\ee
where:
\be
T^{\mu\nu}(x) \equiv {1\over 2} f_{abc} : u_{a} u_{b} F_{c}^{\mu\nu}:
\ee

Let us note the antisymmetry property:
\be
T^{\mu\nu}(x) = - T^{\nu\mu}(x)
\ee
and the analogue of (\ref{gauge-prime}):
\be
d_{Q} T^{\mu\nu}(x) = 0.
\label{gauge-second}
\ee

We also have:
\be
gh(T^{\mu\nu}(x)) = 2.
\ee

The second change is the disappearance of charge conjugation invariance.
Because of these changes we will not be able to prove the disappearance of the
anomalies in higher orders of perturbation theory. Instead, we will be able to
give the generic structure of these anomalies. The computations are similar to
those from the preceding Section but are more complicated from the
combinatorial point of view. Because there are no essential new subtleties we
will give only the results.

Like in the case of QED we write the formul\ae~ (\ref{gauge-inf-1}),
(\ref{gauge-prime-ym}) and (\ref{gauge-second}) in a compact way as follows.
One denotes by
$A^{k}(x),~ k = 1,\dots,11$
the expressions
$T(x),~T^{\mu}(x),~T^{\mu\nu}$;
the index $i$
can take the values
$L, \quad \mu, \quad \mu\nu$
according to the identifications
$
A^{L}(x) \equiv T(x), \quad A^{\mu}(x) \equiv T^{\mu}(x), \quad
A^{\mu\nu}(x) \equiv T^{\mu\nu}(x).
$
Then we can write the gauge invariance conditions in the form
(\ref{gauge-inf-comp-1}):
\be
d_{Q} A^{k}(x) = i \sum_{m} c^{k;\mu}_{m}
{\partial\over \partial x^{\mu}} A^{m}(x), \quad k = 1, \dots, 11
\label{gauge-inf-comp-1-ym}
\ee
for some constants
$c^{k;\mu}_{m}$;
the explicit expressions are:
\be
c^{L;\mu}_{\nu} \equiv \delta^{\mu}_{\nu}, \quad
c^{\nu;\mu}_{\rho\sigma} \equiv {1\over 2} 
\left( \delta^{\nu}_{\rho} \delta^{\mu}_{\sigma}
- \delta^{\mu}_{\rho} \delta^{\nu}_{\sigma} \right)
\label{const-YM}
\ee
and the others are zero. Then we conjecture the following result: one can chose
the chronological products such that, beside the fulfilment of the Bogoliubov
axioms, the following identities are verified:
\be
d_{Q} T(A^{k_{1}}(x_{1}),\dots,A^{k_{p}}(x_{p})) = 
i \sum_{l=1}^{p} (-1)^{s_{l}} \sum_{m} c^{k_{l};\mu}_{m}
{\partial\over \partial x^{\mu}_{l}}
T(A^{k_{1}}(x_{1}),\dots,A^{m}(x_{l}),\dots,A^{k_{p}}(x_{p}))
\label{gauge-inf-comp-ym}
\ee
for all 
$p \in \N$
and all
$k_{1}, \dots, k_{p} = 1,\dots,11$.
Here the expression
$s_{l}$
has the same significance as in the case of QED.

There are a number of facts which can be proved identically. First one can
prove by induction that one can choose the chronological products such that one
has (\ref{Gh}), the symmetry property (\ref{S}) and
\be
T(T^{\mu\nu}(x_{1}),A_{2}(x_{2}),\dots,A_{n}(x_{n})) = 
- T(T^{\nu\mu}(x_{1}),A_{2}(x_{2}),\dots,A_{n}(x_{n})).
\label{s'}
\ee

Next, we can prove that the chronological product can be chosen to depend on
the following fields:
are build only from the fields:
\be
A^{\mu}_{a}, F^{\mu\nu}_{a}, u_{a}, \tilde{u}_{a}, 
\partial_{\mu}\tilde{u}_{a}, \Phi_{a}, \partial_{\mu}\Phi_{a},
\psi_{A}, \overline{\psi}_{A}.
\label{fields}
\ee

Suppose that we have proved the identity (\ref{gauge-inf-comp}) up to the order
$n-1$; then in order in order $n$ we must have a relation of the type
(\ref{gauge-inf-comp-n})
where
$P_{\dots}(X) \equiv P_{\dots}(x_{1},\dots,x_{n})$
are the anomalies having the structure (\ref{wickP}). The maximal degree of the
anomaly is also restricted by (\ref{degP}) and we still have the constraint
(\ref{Gh'}) coming from the ghost number counting. The anomalies will depend on
the following set of fields:
\be
A^{\mu}_{a}, \partial_{\mu}A^{\nu}_{a}, \partial_{\rho}F^{\mu\nu}_{a}, 
u_{a}, \partial_{\mu}u_{a}, \tilde{u}_{a}, 
\partial_{\mu}\tilde{u}_{a}, \partial_{\mu}\partial_{\nu}\tilde{u}_{a},
\Phi_{a}, \partial_{\mu}\Phi_{a}, \partial_{\mu}\partial_{\nu}\Phi_{a},
\psi_{A}, \partial_{\mu}\psi_{A}, \overline{\psi}_{A}, 
\partial_{\mu}\overline{\psi}_{a}
\label{ymC'}
\ee
and the factor
$\partial_{\mu}u_{a}$
can appear only once in any Wick term of the anomaly. Finally, the anomalies 
can be chosen
$SL(2,\C)$-covariant.

From the restrictions (\ref{degP}) and (\ref{Gh'}) we obtain that the possible
anomalies can appear in the following relations:
\bea
d_{Q} T(T(x_{1}),\dots,T(x_{n})) =
\nonumber \\
i \sum_{l=1}^{n} {\partial\over \partial x^{\mu}_{l}} 
T(T(x_{1}),\dots,T^{\mu}(x_{l}),\dots,T(x_{n}))
+ P_{1}(x_{1},\dots,x_{n})
\label{ym1}
\eea
\bea
d_{Q} T(T^{\mu}(x_{1}),T(x_{2}),\dots,T(x_{n})) =
i {\partial\over \partial x^{\mu}_{1}} 
T(T^{\mu\nu}(x_{1}),T(x_{2}),\dots,T(x_{n}))
\nonumber \\
-i \sum_{l=2}^{n} {\partial\over \partial x^{\nu}_{l}} 
T(T^{\mu}(x_{1}),T(x_{2}),\dots,T^{\nu}(x_{l}),\dots,T(x_{n}))
+ P^{\mu}_{2}(x_{1},\dots,x_{n})
\label{ym2}
\eea
\bea
d_{Q} T(T^{\mu}(x_{1}),T^{\nu}(x_{2}),T(x_{3}),\dots,T(x_{n})) =
\nonumber \\
i {\partial\over \partial x^{\rho}_{1}} 
T(T^{\mu\rho}(x_{1}),T^{\nu}(x_{2}),T(x_{3}),\dots,T(x_{n}))
- i {\partial\over \partial x^{\rho}_{2}} 
T(T^{\mu}(x_{1}),T^{\nu\rho}(x_{2}),T(x_{3}),\dots,T(x_{n}))
\nonumber \\
+ i \sum_{l=3}^{n} {\partial\over \partial x^{\rho}_{l}} 
T(T^{\mu}(x_{1}),T^{\nu}(x_{2}),T(x_{3}),\dots,T^{\rho}(x_{l}),\dots,
T(x_{n})) 
+ P^{\mu\nu}_{3}(x_{1},\dots,x_{n}) \qquad
\label{ym3}
\eea
\bea
d_{Q} T(T^{\mu\nu}(x_{1}),T(x_{2}),\dots,T(x_{n})) =
\nonumber \\
i \sum_{l=2}^{n} {\partial\over \partial x^{\rho}_{l}} 
T(T^{\mu\nu}(x_{1}),T(x_{2}),\dots,T^{\rho}(x_{l}),\dots,T(x_{n}))
+ P^{\mu\nu}_{4}(x_{1},\dots,x_{n})
\label{ym4}
\eea
\bea
d_{Q} T(T^{\mu\nu}(x_{1}),T^{\rho}(x_{2}),T(x_{3}),\dots,T(x_{n})) =
\nonumber \\
i {\partial\over \partial x^{\sigma}_{2}} 
T(T^{\mu\nu}(x_{1}),T^{\rho\sigma}(x_{2}),T(x_{3}),\dots,T(x_{n}))
\nonumber \\
- i \sum_{l=3}^{n} {\partial\over \partial x^{\sigma}_{l}} 
T(T^{\mu\nu}(x_{1}),T^{\rho}(x_{2}),\dots,T^{\sigma}(x_{l}),\dots,T(x_{n}))
+ P^{\mu\nu\rho}_{5}(x_{1},\dots,x_{n})
\label{ym5}
\eea
\bea
d_{Q} T(T^{\mu}(x_{1}),T^{\nu}(x_{2}),T^{\rho}(x_{3}),T(x_{4}),\dots,T(x_{n}))
= \nonumber \\
i {\partial\over \partial x^{\sigma}_{1}} 
T(T^{\mu\sigma}(x_{1}),T^{\nu}(x_{2}),T^{\rho}(x_{3}),T(x_{4}),\dots,T(x_{n}))
\nonumber \\
- i {\partial\over \partial x^{\sigma}_{2}} 
T(T^{\mu}(x_{1}),T^{\nu\sigma}(x_{2}),T^{\rho}(x_{3}),T(x_{4}),\dots,T(x_{n}))
\nonumber \\
+ i {\partial\over \partial x^{\sigma}_{3}} 
T(T^{\mu}(x_{1}),T^{\nu}(x_{2}),T^{\rho\sigma}(x_{3}),T(x_{4}),\dots,T(x_{n}))
\nonumber \\
- i \sum_{l=4}^{n} {\partial\over \partial x^{\sigma}_{l}} 
T(T^{\mu}(x_{1}),T^{\nu}(x_{2}),T^{\rho}(x_{3}),T(x_{4}),\dots,
T^{\sigma}(x_{l}),\dots,T(x_{n})) 
\nonumber \\
+ P^{\mu\nu\rho}_{6}(x_{1},\dots,x_{n})
\label{ym6}
\eea
\bea
d_{Q} T(T^{\mu\nu}(x_{1}),T^{\rho\sigma}(x_{2}),T(x_{3}),\dots,T(x_{n})) =
\nonumber \\
i \sum_{l=3}^{n} {\partial\over \partial x^{\lambda}_{l}} 
T(T^{\mu\nu}(x_{1}),T^{\rho\sigma}(x_{2}),T(x_{3}),\dots,T^{\lambda}(x_{l}),
\dots,T(x_{n}))
\nonumber \\
+ P^{\mu\nu\rho\sigma}_{7}(x_{1},\dots,x_{n})
\label{ym7}
\eea
\bea
d_{Q} T(T^{\mu\nu}(x_{1}),T^{\rho}(x_{2}),T^{\sigma}(x_{3}),T(x_{4}),
\dots,T(x_{n})) =
\nonumber \\
i {\partial\over \partial x^{\lambda}_{2}} 
T(T^{\mu\nu}(x_{1}),T^{\rho\lambda}(x_{2}),T^{\sigma}(x_{3}),T(x_{4}),
\dots,T(x_{n}))
\nonumber \\
- i {\partial\over \partial x^{\lambda}_{3}} 
T(T^{\mu\nu}(x_{1}),T^{\rho}(x_{2}),T^{\sigma\lambda}(x_{3}),T(x_{4}),
\dots,T(x_{n}))
\nonumber \\ 
+ i \sum_{l=4}^{n} {\partial\over \partial x^{\lambda}_{l}} 
T(T^{\mu\nu}(x_{1}),T^{\rho}(x_{2}),T^{\sigma}(x_{3}),T(x_{4}),\dots,
T^{\lambda}(x_{l}),\dots,T(x_{n}))
\nonumber \\
+ P^{\mu\nu\rho\sigma}_{8}(x_{1},\dots,x_{n})
\label{ym8}
\eea
\bea
d_{Q} T(T^{\mu}(x_{1}),T^{\nu}(x_{2}),T^{\rho}(x_{3}),T^{\sigma}(x_{4}),\dots,
T(x_{n})) =
\nonumber \\
i {\partial\over \partial x^{\lambda}_{1}} 
T(T^{\mu\lambda}(x_{1}),T^{\nu}(x_{2}),T^{\rho}(x_{3}),T^{\sigma}(x_{4}),
T(x_{5}),\dots,T(x_{n}))
\nonumber \\
- i {\partial\over \partial x^{\lambda}_{2}} 
T(T^{\mu}(x_{1}),T^{\nu\lambda}(x_{2}),T^{\rho}(x_{3}),T^{\sigma}(x_{4}),
T(x_{5}),\dots,T(x_{n}))
\nonumber \\
+ i {\partial\over \partial x^{\lambda}_{3}} 
T(T^{\mu}(x_{1}),T^{\nu}(x_{2}),T^{\rho\lambda}(x_{3}),T^{\sigma}(x_{4}),
T(x_{5}),\dots,T(x_{n}))
\nonumber \\
- i {\partial\over \partial x^{\lambda}_{4}} 
T(T^{\mu}(x_{1}),T^{\nu}(x_{2}),T^{\rho}(x_{3}),T^{\sigma\lambda}(x_{4}),
T(x_{5}),\dots,T(x_{n}))
\nonumber \\
+ i \sum_{l=5}^{n} {\partial\over \partial x^{\lambda}_{l}} 
T(T^{\mu}(x_{1}),T^{\nu}(x_{2}),T^{\rho}(x_{3}),T^{\sigma}(x_{4}),
T(x_{5}),\dots,T^{\lambda}(x_{l}),\dots,T(x_{n})) 
\nonumber \\
+ P^{\mu\nu\rho\lambda}_{9}(x_{1},\dots,x_{n})
\label{ym9}
\eea
where we can assume that:
\bea
P^{\mu\nu}_{3}(X) = 0, \quad P^{\mu\nu\rho}_{5} = 0, \quad
P^{\mu\nu\rho\sigma}_{7} = 0, \quad |X| = 1,
\nonumber \\
P^{\mu\nu\rho}_{6}(X) = 0, \quad P^{\mu\nu\rho\sigma}_{8} = 0,
\quad |X| \leq 2,
\nonumber \\
P^{\mu\nu\rho\sigma}_{9}(X) = 0, \quad |X| \leq 3
\eea
without losing generality. 

From (\ref{S}), we get the following symmetry properties:
\be
P_{1}(x_{1},\dots,x_{n}) \quad
{\rm is~ symmetric~ in} \quad x_{1},\dots,x_{n};
\label{s1}
\ee
\be
P_{2}^{\mu}(x_{1},\dots,x_{n}) \quad
{\rm is~ symmetric~ in} \quad x_{2},\dots,x_{n};
\label{s2}
\ee
\be
P_{3}^{\mu\nu}(x_{1},\dots,x_{n}) \quad
{\rm is~ symmetric~ in} \quad x_{3},\dots,x_{n};
\label{s3}
\ee
\be
P_{4}^{\mu\nu}(x_{1},\dots,x_{n}) \quad
{\rm is~ symmetric~ in} \quad x_{2},\dots,x_{n};
\label{s4}
\ee
\be
P_{5}^{\mu\nu\rho}(x_{1},\dots,x_{n}) \quad
{\rm is~ symmetric~ in} \quad x_{3},\dots,x_{n};
\label{s5}
\ee
\be
P_{6}^{\mu\nu\rho}(x_{1},\dots,x_{n}) \quad
{\rm is~ symmetric~ in} \quad x_{4},\dots,x_{n};
\label{s6}
\ee
\be
P_{7}^{\mu\nu\rho\sigma}(x_{1},\dots,x_{n}) \quad
{\rm is~ symmetric~ in} \quad x_{3},\dots,x_{n};
\label{s7}
\ee
\be
P_{8}^{\mu\nu\rho\sigma}(x_{1},\dots,x_{n}) \quad
{\rm is~ symmetric~ in} \quad x_{4},\dots,x_{n};
\label{s8}
\ee
\be
P_{9}^{\mu\nu\rho\sigma}(x_{1},\dots,x_{n}) \quad
{\rm is~ symmetric~ in} \quad x_{5},\dots,x_{n};
\label{s9}
\ee
we also have:
\be
P_{3}^{\mu\nu}(x_{1},\dots,x_{n}) \quad
{\rm is~ antisymmetric~ in} \quad (x_{1},\mu), (x_{2},\nu);
\label{s3'}
\ee
\be
P_{4}^{\mu\nu} = - P_{4}^{\nu\mu};
\label{s4'}
\ee
\be
P_{5}^{\mu\nu\rho} = - P_{5}^{\nu\mu\rho};
\label{s5'}
\ee
\be
P_{6}^{\mu\nu\rho}(x_{1},\dots,x_{n}) \quad
{\rm is~ antisymmetric~ in} \quad (x_{1},\mu), (x_{2},\nu), (x_{3},\rho);
\label{s6'}
\ee
\be
P_{7}^{\mu\nu\rho\sigma} = - P_{7}^{\nu\mu\rho\sigma}
= - P_{7}^{\mu\nu\sigma\rho};
\label{s7'}
\ee
\be
P_{7}^{\mu\nu\rho\sigma}(x_{1},x_{2},\dots,x_{n}) =
P_{7}^{\rho\sigma\mu\nu}(x_{2},x_{1},\dots,x_{n});
\label{s7"}
\ee
\be
P_{8}^{\mu\nu\rho\sigma} = - P_{8}^{\nu\mu\rho\sigma};
\label{s8'}
\ee
\be
P_{8}^{\mu\nu\rho\sigma}(x_{1},x_{2},x_{3},\dots,x_{n}) =
- P_{8}^{\mu\nu\sigma\rho}(x_{1},x_{3},x_{2},\dots,x_{n});
\label{s8"}
\ee
\be
P_{9}^{\mu\nu\rho\sigma}(x_{1},\dots,x_{n}) \quad
{\rm is~ antisymmetric~ in} \quad (x_{1},\mu), (x_{2},\nu), (x_{3},\rho),
(x_{4},\sigma).
\label{s9'}
\ee

Let us note that for
$n = 2$
only the first five relations (\ref{ym1})-(\ref{ym5}) have to be checked; this
can be done by some long but straightforward computations.

\subsection{The Generic Structure of the Anomalies}

If we apply the operator
$d_{Q}$
to the anomalous relations (\ref{ym1})-(\ref{ym9}) we easily obtain again
consistency relations of the Wess-Zumino type:
\be
d_{Q} P_{1}(x_{1},\dots,(x_{n}) 
= i \sum_{l=1}^{n} {\partial\over \partial x^{\mu}_{l}} 
P^{\mu}_{2}(x_{l},x_{1},\dots,\hat{x}_{l},\dots,x_{n})
\label{ymG1}
\ee
\be
d_{Q} P^{\mu}_{2}(x_{1}),\dots,(x_{n})
= i {\partial\over \partial x^{\nu}_{1}}
P^{\mu\nu}_{4}(x_{1},\dots,x_{n})
-i \sum_{l=2}^{n} {\partial\over \partial x^{\nu}_{l}} 
P^{\mu\nu}_{3}(x_{1},x_{l},x_{2},\dots,\hat{x}_{l},\dots,x_{n})
\label{ymG2}
\ee
\bea
d_{Q} P^{\mu\nu}_{3}(x_{1},\dots,x_{n})
= i {\partial\over \partial x^{\rho}_{1}}
P^{\mu\rho\nu}_{5}(x_{1},\dots,x_{n})
- i {\partial\over \partial x^{\rho}_{2}}
P^{\nu\rho\mu}_{5}(x_{2},x_{1},x_{3},\dots,x_{n})
\nonumber \\
+ i \sum_{l=3}^{n} {\partial\over \partial x^{\rho}_{l}}
P^{\mu\nu\rho}_{4}(x_{1},x_{2},x_{l},x_{3},\dots,\hat{x}_{l},\dots,x_{n})
\label{ymG3}
\eea
\be
d_{Q} P^{\mu\nu}_{4}(x_{1},x_{n})
= i \sum_{l=2}^{n} {\partial\over \partial x^{\rho}_{l}} 
P^{\mu\nu\rho}_{5}(x_{1},x_{l},x_{2},\dots,\hat{x}_{l},\dots,x_{n})
\label{ymG4}
\ee
\bea
d_{Q} P^{\mu\nu\rho}_{5}(x_{1},\dots,x_{n})
= i {\partial\over \partial x^{\sigma}_{2}}
P^{\mu\nu\rho\sigma}_{7}(x_{1},\dots,x_{n})
\nonumber \\
- i \sum_{l=3}^{n} {\partial\over \partial x^{\sigma}_{l}}
P^{\mu\nu\rho\sigma}_{8}(x_{1},x_{2},x_{l},x_{3},\dots,\hat{x}_{l},\dots,x_{n})
\label{ymG5}
\eea
\bea
d_{Q} P^{\mu\nu\rho}_{6}(x_{1},\dots,x_{n})
= i {\partial\over \partial x^{\sigma}_{1}}
P^{\mu\sigma\nu\rho}_{8}(x_{1},\dots,x_{n})
\nonumber \\
- i {\partial\over \partial x^{\sigma}_{2}}
P^{\nu\sigma\mu\rho}_{8}(x_{2},x_{1},x_{3},\dots,x_{n})
+ i {\partial\over \partial x^{\sigma}_{3}}
P^{\rho\sigma\mu\nu}_{8}(x_{3},x_{2},x_{1},x_{4},\dots,x_{n})
\nonumber \\
- i \sum_{l=4}^{n} {\partial\over \partial x^{\rho}_{l}}
P^{\mu\nu\rho\sigma}_{9}(x_{1},x_{2},x_{3},x_{l},x_{4},\dots,\hat{x}_{l},
\dots,x_{n})
\label{ymG6}
\eea
\be
d_{Q} P^{\mu\nu\rho\sigma}_{i}(x_{1},\dots,x_{n}) = 0, \quad i = 7,8,9.
\label{ymG789}
\ee

After a long computation (using the symmetry properties, the ghost number
restrictions, etc. and making some convenient finite renormalizations) one can
determine the generic form of the anomalies.  One starts from a generic form of
the same type as in the case of QED for all anomalies
$P^{\dots}_{i}, \quad i = 1,\dots,9$
and determines that:
\be
P^{\dots}_{i} = 0, \quad i = 3,\dots,9
\ee
and
$P^{\dots}_{i} = 0, \quad i = 1,2$
can be chosen of the form:
\be
P_{1} = \delta(X) W_{1}(x_{1}), \quad 
P^{\mu}_{2} = \delta(X) W^{\mu}_{2}(x_{1}).
\ee

The gauge invariance condition reduces to:
\be
d_{Q} W_{1} = i \partial_{\mu} W^{\mu}_{2}.
\ee

We give now the generic form of the Wick polynomials
$W_{i}, \quad i = 1,2$.
fulfilling these conditions. First we have:
\bea
W^{\mu}_{2} = c_{abcd} :u_{a} u_{b} \Phi_{c} \partial^{\mu}\Phi_{d}:
+ c_{abc} :u_{a} u_{b} \partial^{\mu}\Phi_{c}:
\nonumber \\
+ c_{ab;AB} :u_{a} u_{b} \overline{\psi}_{A}\gamma^{\mu}\psi_{B}:
+ c_{ab;AB}' :u_{a} u_{b} \overline{\psi}_{A}\gamma^{\mu}\gamma_{5}\psi_{B}:
\eea
where:
\bea
c_{abcd} = - (c \leftrightarrow d) = - (a \leftrightarrow b), \quad
c_{ab;AB} = - (a \leftrightarrow b), \quad c_{ab;AB}' = - (a \leftrightarrow b)
\nonumber \\
c_{abcd} = 0 \quad
{\rm iff} \quad m_{a} + m_{b} + m_{c} + m_{d} \geq 0
\nonumber \\
c_{abc} = 0, \quad m_{a} + m_{b} + m_{c} \geq 0, \quad
c_{ab;AB} = 0, \quad c_{ab;AB}' = 0, \quad
{\rm iff} \quad m_{a} + m_{b} \geq 0.
\eea

Finally we have:
\bea
W_{1} = 
- 2 c_{abcd} :u_{a} A^{\rho}_{b} \Phi_{c} \partial_{\rho}\Phi_{d}:
- 2 c_{abc} :u_{a} A^{\rho}_{b} \partial_{\rho}\Phi_{c}:
\nonumber \\
- 2 c_{ab;AB} :u_{a} A^{\rho}_{b} \overline{\psi}_{A}\gamma_{\rho}\psi_{B}:
- 2 c_{ab;AB}' :u_{a} A^{\rho}_{b} 
\overline{\psi}_{A}\gamma_{\rho}\gamma_{5}\psi_{B}:
\nonumber \\
+ d'_{abc} \left( :u_{a} \partial_{\rho}\Phi_{b} \partial^{\rho}\Phi_{c}:
+ m_{b} m_{c} :u_{a} A_{b}^{\rho} A_{c\rho}:
- 2 m_{b} :u_{a} A^{\rho}_{b} \partial_{\rho}\Phi_{c}: \right)
\nonumber \\
+ d_{abc} :u_{a}\Phi_{b} \Phi_{c}:
+ d_{abcd} :u_{a}\Phi_{b} \Phi_{c} \Phi_{d}:
+ d_{abcde} :u_{a}\Phi_{b} \Phi_{c} \Phi_{d} \Phi_{e}:
\nonumber \\
+ d_{ab;AB} :u_{a} \overline{\psi}_{A} \psi_{B}:
+ d_{ab;AB}' :u_{a} \overline{\psi}_{A}\gamma_{5}\psi_{B}:
\nonumber \\
+ f_{abc} :u_{a} F_{b}^{\rho\sigma} F_{c\rho\sigma}:
+ f'_{abc} \varepsilon_{\mu\nu\rho\sigma} 
:u_{a} F_{b}^{\mu\nu} F^{\rho\sigma}_{c}:
\eea
where:
\bea
d_{abc} = 0 \quad m_{a} + m_{b} + m_{c} > 0, \quad 
d_{abcd} = 0 \quad m_{a} + m_{b} + m_{c} + m_{d} > 0, \quad
\nonumber \\
d_{abcde} = 0, \quad m_{a} + m_{b} + m_{c} + m_{d} + m_{e} > 0,
\nonumber \\
f_{abc} = 0, \quad f'_{abc} = 0, \quad d'_{abc} = 0, \quad 
\quad m_{a} > 0,
\nonumber \\
d_{ab;AB} = 0, \quad d_{ab;AB}' = 0,
\quad m_{a} + m_{b} > 0.
\eea

There are no obvious arguments for the elimination of these anomalies. We
remark a very interesting fact: if all the Bosons are heavy, then there the
expression of the anomalies simplifies considerably. 

We close with another interesting remark. Let us define the following
differential forms:
\be
{\cal T}_{p}(X) \equiv \sum T(A^{k_{1}}(x_{1}),\dots,A^{k_{p}}(x_{p}))
dx_{1;k_{1}} \wedge \dots \wedge dx_{p;k_{p}}
\label{comp-temp}
\ee
where we have defined in general:
\be
dx_{L} \equiv dx \equiv dx^{0} \wedge \dots \wedge dx^{3}, \quad
dx_{\mu} \equiv i_{\partial^{\mu}}dx, \quad
dx_{\rho\sigma} \equiv i_{\partial^{\rho}}i_{\partial^{\sigma}}dx.
\ee

It is a very interesting fact that the following relation is true:
\be
d^{\rho} \wedge dx_{i} = \sum_{j} c^{j;\rho}_{i} dx_{j}
\label{dx}
\ee
where the constants
$c^{j;\rho}_{i}$
are exactly the same as those appearing in (\ref{const-YM}). Then it is easy to
prove that the induction hypothesis can be compactly written as
\be
d_{Q} {\cal T}_{p}(X) = i d {\cal T}_{p}(X), \quad p = 1,\dots, n-1
\ee
and the anomalous gauge identity in order $n$ is:
\be
d_{Q} {\cal T}_{n}(X) = i d {\cal T}_{n}(X) + {\cal P}_{n}(X);
\label{comp-TP}
\ee
here the anomaly
${\cal P}_{n}(X)$
has an expression of the type (\ref{comp-temp}):
\be
{\cal P}_{p}(X) \equiv \sum P^{k_{1},\dots,k_{p}}(X)
dx_{1;k_{1}} \wedge \dots \wedge dx_{p;k_{p}}
\label{comp-temp-ano}
\ee
with the identifications:
\bea
P^{L,\dots,L} = P_{1}, \quad
P^{\mu,L,\dots,L} = P^{\mu}_{2}, \quad
P^{\mu,\nu,L,\dots,L} = P^{\mu\nu}_{3}, \quad
\nonumber \\
P^{\mu\nu,L,\dots,L} = P^{\mu\nu}_{4}, \quad
P^{\mu\nu,\rho,L,\dots,L} = P^{\mu\nu\rho}_{5}
P^{\mu,\nu,\rho,L,\dots,L} = P^{\mu\nu\rho}_{6}, \quad
\nonumber \\
P^{\mu\nu,\rho\sigma,L,\dots,L} = P^{\mu\nu\rho\sigma}_{7}, \quad
P^{\mu\nu,\rho,\sigma,L,\dots,L} = P^{\mu\nu\rho\sigma}_{8}, \quad
P^{\mu,\nu,\rho,\sigma,L,\dots,L} = P^{\mu\nu\rho\sigma}_{9}.
\eea

So, the expressions
${\cal P}_{p}(X)$
are differential forms with coefficients quasi-local operators. Let us denote
by 
$\cal A$
this class of differential forms.  From (\ref{comp-TP}) we easily obtain the
consistency equation
\be
d_{Q} {\cal P}_{n}(X) + i d {\cal P}_{n}(X) = 0
\ee
which is the compact form of the relations (\ref{ymG1}) - (\ref{ymG789}). One
can ``solve" this equation using the homotopy operator $p$ of the de Rham
complex: we have
\be
{\cal P}_{n}(X) =  d(p {\cal P}_{n}(X)) + d_{Q}(i p {\cal P}_{n}(X)).
\ee

It is tempting to argue that by the finite renormalization
\be
{\cal T}_{n}(X) \rightarrow {\cal T}_{n}(X) + i p {\cal P}_{n}(X)
\ee
the anomalies are eliminated. However one can check that if we apply the
homotopy operator $p$ on a element from
$\cal A$
we do not obtain a element from
$\cal A$.
It follows that the finite renormalization given above is not legitimate and
the argument has to be modified somehow. However, let us notice the interesting
fact that the usual expression of the homotopy operator for the de Rham complex
is constructed using the action of the dilation group. This is in agreement to
the role played by this group in the traditional approach to the
non-renormalizability theorems.

\end{document}